\documentclass[aps,prapplied,twocolumn,superscriptaddress]{revtex4-2}

\usepackage{graphicx}
\usepackage{amsmath,amssymb}
\usepackage{hyperref}
\usepackage{xcolor}
\usepackage{float}

\def\scititle{
 	Cavity QED with molecular defects coupled to a photonic crystal cavity
 } 
\title{\bfseries \boldmath \scititle}

\newcommand{\Chem}{Department of Chemistry, Purdue University, West Lafayette, IN 47907, USA}
\newcommand{\Phys}{Department of Physics and Astronomy, Purdue University, West Lafayette, IN 47907, USA}
\newcommand{\ECE}{Department of Electrical and Computer Engineering, Purdue University, West Lafayette, IN 47907, USA}

\begin{document}

\title{\scititle} 

\author{Christian~M.~Lange}
\affiliation{\Phys}

\author{Arya~D.~Keni}
\affiliation{\ECE}

\author{Ishita~Agarwal}
\affiliation{\Phys}

\author{Emma~Daggett}
\affiliation{\Chem}

\author{Adhyyan~S.~Mansukhani}
\affiliation{\Phys}

\author{Ankit~Kundu}
\affiliation{\ECE}

\author{Benjamin~Cerjan}
\affiliation{\Chem}

\author{Libai~Huang}
\affiliation{\Chem}

\author{Jonathan~D.~Hood}
\email{hoodjd@purdue.edu}
\affiliation{\Chem}
\affiliation{\Phys}



\begin{abstract} \bfseries \boldmath
We implement permanent spectral tuning to bring lifetime-limited emitters into collective resonance within an integrated photonic cavity. This addresses a fundamental challenge in solid-state cavity QED: combining multiple coherent quantum emitters with scalable nanophotonics. Our hybrid approach decouples emitter synthesis from nanophotonic fabrication using straightforward techniques that make cavity QED broadly accessible. High doping densities allow us to couple several coherent emitters to a single cavity mode, while optically-induced frequency shifting provides long-lived spectral control. By tuning two molecules into resonance, we demonstrate controlled formation of collective quantum states, establishing a scalable platform for many-body cavity QED. This opens pathways toward chemically-designed quantum systems where optical properties are engineered through synthetic chemistry. 
\end{abstract}

\maketitle

Cavity quantum electrodynamics (cavity QED) was originally pioneered using ultracold atoms in free-space optical cavities~\cite{thompson1992observation, kuhn2002deterministic, reiserer2015cavity}, establishing the experimental foundation for quantum networks~\cite{kimble2008quantum}, single-photon nonlinearities~\cite{chang2014quantum}, and photon generation~\cite{kuhn2002deterministic,kuzmich2003generation, thomas2022efficient}. Although these atomic systems demonstrated exquisite quantum control, they faced fundamental scaling limitations, prompting a shift toward solid-state platforms with integrated nanophotonic cavities~\cite{lodahl2015interfacing, englund2010deterministic, faraon2012coupling}. Chip-based systems promise miniaturized and scalable quantum technologies, but the transition to solid-state systems comes at a significant cost: the optical coherence of most solid-state emitters is compromised through interactions with their host lattices. Environmental interactions broaden linewidths beyond the lifetime limit as states decohere faster than they decay, while environmental differences between emitters cause transition frequencies to be separated by much more than the transition linewidths. 

As a result of inhomogeneous broadening and decoherence, the realization of collective quantum effects—where multiple emitters interact coherently through a shared electromagnetic field—has remained largely confined to ultracold atomic systems~\cite{yan2023superradiant, ritter2012elementary} and a few recent solid-state platforms where emitters can be precisely tuned into resonance~\cite{trebbia2022tailoring, rattenbacher2023onchip,lange2024superradiant, nobakht2024cavity, lukin2023twoemitter}. 
Despite significant advances in single emitter coupling, the field still lacks an accessible cavity QED platform that combines coherent lifetime-limited emitters, strong light-matter coupling, and scalability to multiple interacting emitters.

\begin{figure*}[tb!]
	\centering
	\includegraphics[width=0.9\textwidth]{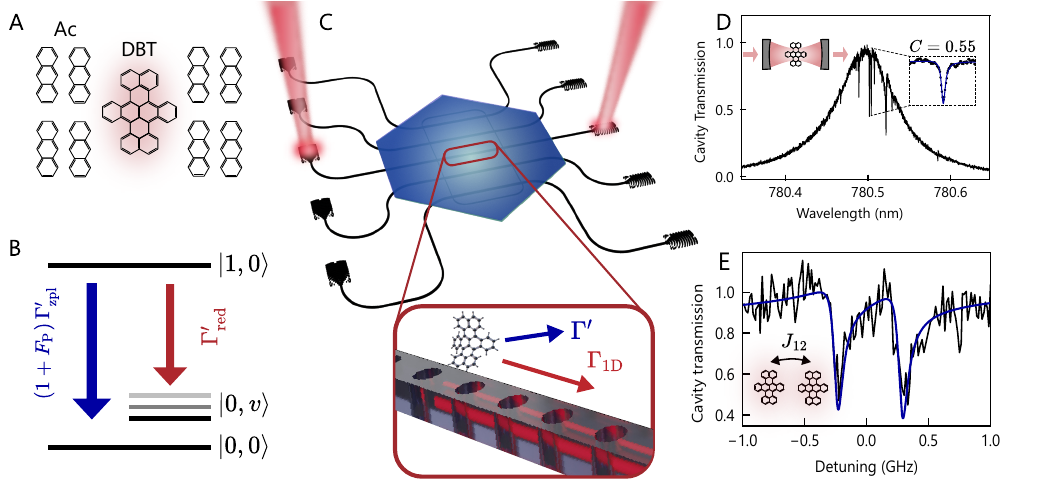} 
	\caption{\textbf{Integration of DBT molecules with photonic crystal cavities.}
    \textbf{(A)} Dibenzoterrylene (DBT) molecule doped into an anthracene (Ac) crystal matrix. 
    \textbf{(B)} Energy level diagram of DBT. In a bulk crystal at 3~K, the linewidth of the transition is 40 MHz, limited by the $\sim 4$ ns lifetime. The 0-0 zero-phonon line (shown in blue) comprises 30\% of the decay, while the rest goes to vibrational states and their phonon sidebands (shown in red). In a cavity, the 0-0 zero-phonon line is enhanced by the Purcell factor $F_\mathrm{P}$. 
    \textbf{(C)} Nanophotonic integration of DBT. A 200 nm thick crystal of anthracene is doped with DBT and stamped onto a photonic crystal cavity. The cavity transmission is probed with grating couplers. $\Gamma_{\mathrm{1D}}$ and $\Gamma^\prime$ denote the decay rate into the cavity vs. free-space. 
    \textbf{(D)} Transmission spectrum of the photonic crystal cavity. Cavity-molecule coupling is manifested as sharp dips. The inset shows a molecule antiresonance with a cooperativity of 0.55.
    \textbf{(E)} Two molecules interacting through the cavity mode in the dispersive regime. 
    }
	\label{fig:fig_1} 
\end{figure*}

In this work, we present a hybrid cavity QED system in which lifetime-limited organic molecules~\cite{tamarat1999ten, nicolet2007single, adhikari2023future} are integrated with photonic crystal cavities, achieving strong light-matter coupling and collective interactions among multiple emitters. Our modular approach decouples emitter synthesis from nanophotonic fabrication, offering excellent coherence, high optical quality nanophotonics,  and ease of implementation.
Dibenzoterrylene (DBT) molecules~\cite{tamarat2000ten, toninelli2021single, chu2017single, grandi2016quantum} are embedded as defects in an anthracene crystal (see Fig.~\ref{fig:fig_1}A). The optically thin, high-purity crystal is stamped onto a photonic crystal cavity (see Fig.~\ref{fig:fig_1}C). Because DBT has a well-known insertion site in anthracene, this micropositioning technique gives precise control of the emitter's dipole moment with respect to the cavity electric field. 

The high density of coherent emitters allows us to couple several molecules to a single cavity mode with cooperativities of $C\approx0.5$ (Fig.~\ref{fig:fig_1}D). We prepare states of two interacting molecules (Fig.~\ref{fig:fig_1}E) in both the resonant (dissipative) and detuned (dispersive) regimes. Using a light-induced frequency shifting mechanism, we permanently tuned pairs of molecules into resonance to form collective systems.

This work establishes a new platform for solid-state cavity QED that combines the exceptional coherence of molecular emitters with the scalability of integrated photonics. By demonstrating coherent coupling of multiple lifetime-limited emitters within a single cavity mode, we open pathways to explore collective quantum phenomena, including superradiance, many-body entanglement, and non-classical light generation~\cite{gonzalez-tudela_deterministic_2015, martin2014large, paulisch2019quantum}. The successful application of molecular emitters in a quantum photonic system points to a future of chemically synthesized qubits, where quantum properties become designable characteristics through established chemical synthesis techniques rather than materials engineering~\cite{wasielewski2020exploiting}. 

Our quantum emitter consists of DBT molecules doped in a high-purity anthracene crystal, as illustrated in Fig.~\ref{fig:fig_1}A. DBT is well-established as an exceptional quantum emitter and test bench of solid-state quantum optics~\cite{toninelli2021single}. As shown in Fig.~\ref{fig:fig_1}B, the level structure of DBT consists of a two-level electronic transition with a lifetime of $\sim 4$ nanoseconds. The zero-phonon line (ZPL) transition accounts for approximately 30\% of the emission~\cite{trebbia2009efficient}, with the remainder coupling to a manifold of vibrational states and their corresponding phonon sidebands~\cite{zirkelbach2022high}. At 3~K, phonon dephasing due to interactions with the anthracene lattice freezes out, resulting in lifetime-limited linewidths of 40 MHz. DBT exhibits other remarkable optical properties: exceptional spectral stability due to both the rigid anthracene host matrix and the lack of a first-order Stark shift~\cite{ren2022photoniccircuited}; brightness stemming from negligible nonradiative decay~\cite{ren2022probing, musavinezhad2023quantum}; and minimal intersystem crossing to the triplet state (yield~$<10^{-7}$)~\cite{nicolet2007single}. These properties make DBT an excellent source of indistinguishable photons, demonstrating 93\% indistinguishability from a single emitter~\cite{rezai2018coherence} and 70\% between separate emitters~\cite{duquennoy2023singular}---both remarkably achieved without the use of an optical cavity. 

\begin{figure*}[tb!] 
	\centering
	\includegraphics[width=0.9\textwidth]{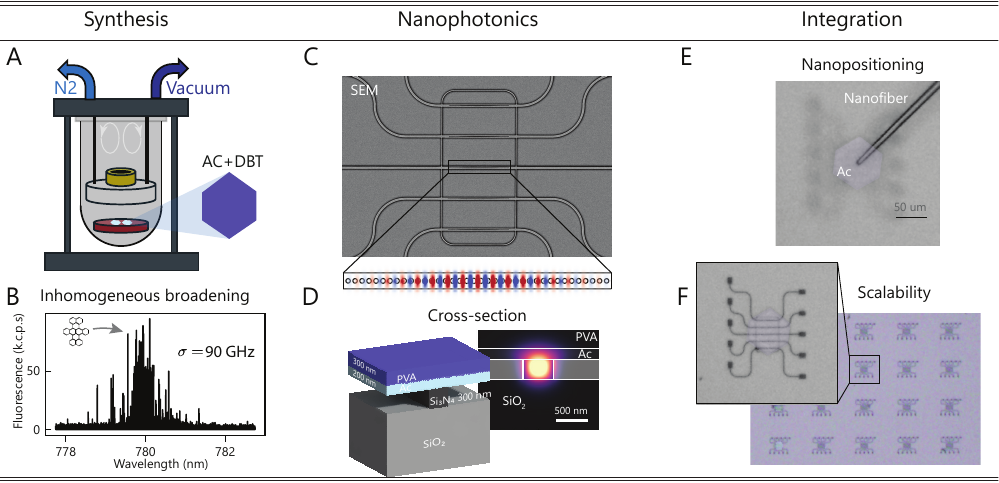} 
\caption{\textbf{Synthesis and micropositioning of DBT-doped anthracene crystals.}
\textbf{(A)} Thin anthracene crystals doped with DBT are synthesized in a homebuilt apparatus under controlled atmosphere.
\textbf{(B)} Fluorescence excitation spectrum of DBT molecules in an anthracene crystal at 3 K, showing individual resonances distributed around 780 nm. The inhomogeneous distribution has a standard deviation of $\sigma = 90$ GHz, while each peak represents a single molecule with a linewidth of approximately 40 MHz.
\textbf{(C)} SEM image of photonic crystal cavities with silicon nitride protective rectangles designed to prevent PVA from contacting the cavity region.
\textbf{(D)} Cross-sectional schematic of the integrated device structure and simulation of the confined optical mode in the silicon nitride photonic crystal cavity. 
\textbf{(E)} A tapered optical fiber is used to precisely position an anthracene crystal (highlighted in blue) onto a photonic structure, with the molecular dipoles naturally aligned to the cavity electric field.
\textbf{(F)} Multiple crystals positioned on an array of cavities, representing hundreds of cavity-coupled systems and demonstrating the excellent scalability of this approach.
}
\label{fig:fig_2} 
\end{figure*}

Coupling an emitter to a nanophotonic cavity serves two roles in enhancing the emission of indistinguishable photons. The Purcell factor enhances the decay rate of the ZPL and also increases the proportion of photons emitted into the guided mode. 

The integration of organic molecules with nanophotonics has been an outstanding challenge. Unlike defect centers in diamond and epitaxial quantum dots, the host materials of emitters like dibenzoterrylene cannot be easily structured into low-loss nanophotonics. Photonic integration with organic molecules therefore requires a hybrid incorporation scheme, such as reflowing DBT in a molten host matrix~\cite{shkarin2021nanoscopic, boissier2021coherent, turschmann2017chipbased, gmeiner2016spectroscopy,rattenbacher2019coherent} and embedding DBT in a polyethylene film \cite{rattenbacher2023onchip}. Both of these integration strategies require the use of a disordered host matrix, which decreases the stability of the molecular transition and the optical quality of the photonic device. A recent demonstration showed that DBT in an optically pure anthracene crystal could be coupled to a nanophotonic waveguide~\cite{ren2022photoniccircuited}.


Fig.~\ref{fig:fig_1}C illustrates our scheme to couple DBT molecules to a nanobeam photonic crystal cavity. We dope DBT into high-purity anthracene crystals with dimensions of 200 nm thickness and 30--50 \textmu m width following protocols established in Refs.~\cite{wei2020singlemolecule, ren2022photoniccircuited}. The crystal is precisely positioned on top of a high-quality silicon nitride photonic crystal cavity designed according to Ref.~\cite{quan2011deterministic}. Notably, the crystalline anthracene preserves the optical quality of the cavity, with the quality factor decreasing by as little as 10\% after integration. As shown in Fig.~\ref{fig:fig_2}D, the cavity's evanescent field extends into the anthracene crystal, enabling coupling between the molecular emitters and the confined optical mode. 

The crystals are synthesized in a homemade co-sublimation setup (Fig.~\ref{fig:fig_2}A). DBT molecules incorporate into the anthracene matrix with minimal structural distortion, allowing high doping concentrations without degrading their optical properties. Our crystal growth method yields a density of approximately 100 DBT molecules per cubic micron, guaranteeing multiple emitters within the cavity region. As shown in Fig.~\ref{fig:fig_2}B, the zero-phonon lines of these molecules are distributed around 780~nm with a standard deviation of $\sigma=90$~GHz. Because of the 40~MHz linewidth of individual molecules, the molecules can be spectrally isolated and addressed even at high densities. 



As shown in Fig.~\ref{fig:fig_2}E-F, the high aspect-ratio crystals are positioned onto photonic crystal cavities using a tapered optical fiber as a micropositioning tool~\cite{ren2022photoniccircuited}. Because the transition dipole moment of DBT is known to align with the crystalline b-axis of anthracene~\cite{nicolet2007single}, this method enables orientation of the molecular dipole moments with the TE$_{00}$-polarized electric field of the cavity, maximizing the light-matter coupling strength. 
%

To prevent sublimation of anthracene at room temperature, we coat the crystals with a 300 nm thick layer of polyvinyl alcohol (PVA). To maintain the cavity's high quality factor, we fabricate a silicon nitride ring (Fig.~\ref{fig:fig_2}C) around each cavity, which prevents the amorphous polymer from contacting the sensitive cavity region. 
Because the presence of anthracene and PVA shifts the cavity resonance as much as 10 THz, the cavities are designed to be blue detuned with respect to the DBT resonances.
We designed cavities with a range of center wavelengths and placed a crystal on five cavities per placement process, increasing the likelihood of a cavity resonance aligning with the molecular resonances. This integration approach is highly scalable, as demonstrated in Fig.~\ref{fig:fig_2}F, where we successfully positioned crystals on an array of 100 cavities on a single photonic chip. 
%
\begin{figure*}[tb!] 
	\centering
 	\includegraphics[width=0.9\textwidth]{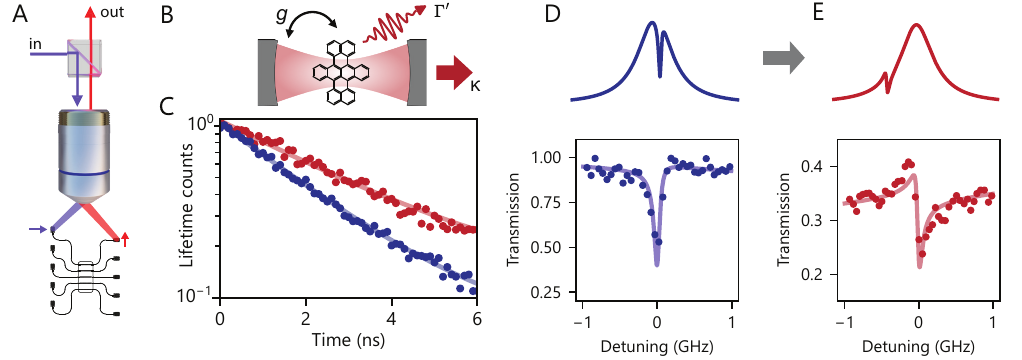} 
        \caption{\textbf{Characterization of DBT-cavity system.}
        \textbf{(A)} Schematic of the optical setup used for lifetime and transmission measurements.
        \textbf{(B)} A molecule is coupled to a cavity with a coupling strength $g$, cavity decay rate $\kappa$, and free-space emission rate $\Gamma'$. 
        \textbf{(C)} Decay rate of Purcell enhanced molecule at a molecule-cavity detuning of $\Delta_{mc}=5.6$~GHz (blue) and $\Delta_{mc}=-31$~GHz (red) giving a free-space decay rate of $\Gamma^\prime=40$ MHz and a cavity coupling strength of $g=0.6$~GHz.
        \textbf{(D)} Experimental cavity transmission spectrum (blue dots) with theoretical fit (blue line) for the molecule at a detuning of $\Delta_{mc}=5.6$~GHz.
        \textbf{(E)} Experimental transmission spectrum (red dots) with theoretical fit (red line) for the same molecule at a detuning of $\Delta_{mc}=-31$~GHz. The spectra in (D) and (E) were fit to give a pure dephasing rate of $\gamma^\ast=10$ MHz, giving cavity QED parameters of $(g,\kappa,\gamma)/2\pi=(0.6, 45, 0.06)$ GHz and a cooperativity of $C=0.53$. 
        }
	\label{fig:fig_3} 
\end{figure*}
%
The cavity-emitter coupling is described by the cavity QED parameters $g$, $\kappa$, and $\gamma$, where $g$ is the emitter-cavity coupling strength, $\kappa$ is the decay rate of the cavity, and $\gamma=\Gamma'+2\gamma^*$ is the decoherence of the emitter. Here, $\Gamma'=\Gamma'_{\text{zpl}} + \Gamma'_{\text{red}}$ is the free space decay rate and $\gamma^*$ is the pure dephasing. 
The enhancement of a transition by the cavity is characterized by the Purcell factor, $F_\mathrm{P}=\Gamma_{\mathrm{1D}}/\Gamma^\prime_{\text{zpl}}$, where $\Gamma_{\mathrm{1D}}$ is the decay rate into the cavity. $\Gamma_{\mathrm{1D}}$ is related to the cavity QED parameters as $\Gamma_{\mathrm{1D}}=\frac{g^{2}\kappa}{(\kappa/2)^{2}+\Delta_{mc}^{2}}$, where $\Delta_{mc}$ is the detuning of the molecule from the cavity. 
For a single emitter coupled to a cavity, the input-output formalism yields the transmission spectrum
~\cite{asenjo2017atom}
\begin{equation}
   T(\omega)=\left|\frac{\kappa/2}{\mathrm{i}\Delta_c-\frac{\kappa}{2}+\frac{g^2}{\mathrm{i}\Delta_0-\gamma/2}}\right|^2
\label{eq:transmission}
\end{equation}
where $\Delta_0=\omega_L-\omega_0$ is the detuning of the emitter frequency $\omega_0$ from the laser frequency $\omega_L$ and $\Delta_c=\omega_L-\omega_c$ is the detuning of the cavity frequency $\omega_c$ from the laser frequency. 
As shown in the transmission spectra of Fig.~\ref{fig:fig_3}D-E,  the coupling to individual  DBT molecules is evidenced by sharp dips in cavity transmission, corresponding to the emitters' resonances. The depth of the emitter antiresonance is a measure of the coupling strength. For a weakly coupled emitter, the depth of the transmission dip is  $\Delta T \approx 2C$, where the cooperativity is defined as $C=\frac{4 g^2}{\kappa\gamma}$. 

The molecules are coupled to a photonic crystal cavity~\cite{quan2011deterministic} with a quality factor of $\sim 10,000$ and mode volume of $2.8\,(\lambda/n)^3$.  We probe the cavity-emitter coupling through the cavity transmission signal. Light is coupled into and out of the waveguides through inverse-designed grating couplers~\cite{tidy3d} ($25\%$ efficiency).  Light is delivered via and collected through an objective above a Montana cryostat at 3~K.  The input and output grating couplers are coupled to single mode fibers and oriented perpendicular to each other so that a polarizing beamsplitter can be used to separate the two beam paths. 

We characterized the cavity QED parameters of a molecule by measuring its lifetime and transmission spectrum at different detunings. 
The transmission spectrum of the bare cavity is fit to a Lorentzian to extract $\kappa/2\pi =45$~GHz, corresponding to a quality factor of $Q=8,600$. The cavity resonance was tuned passively due to slow adsorption of gases in the cryostat~\cite{lukin2023twoemitter}. The cryostat pressure is maintained by cryopumping, and any gas that leaks in will stick to surfaces. In this case, this caused a slow drift of the cavity resonance of tens of GHz over the course of several days. 

At a cavity detuning of $ \Delta_{mc} = -31 $~GHz, the excited state lifetime of the molecule measured 3.13~ns. At a detuning of $ \Delta_{mc} = 5.6 $~GHz, a lifetime of 2.2~ns was measured, allowing us to extract a free-space decay rate of $\Gamma^\prime/2\pi=40$~MHz and a cavity coupling strength of $g=0.6$~GHz. Assuming a branching ratio of the electronic transition to the total free-space decay of $\Gamma^\prime_{\mathrm{zpl}}/\Gamma^\prime\approx 33\%$, this gives a Purcell factor of $F_\mathrm{P}=2.4$. 

Some small proportion of the decay to the phonon sideband is also modified by the presence of the cavity. However, the phonon sideband is centered on the order of THz away from the ZPL~\cite{clear2020phonon} and is a few THz wide. Therefore, the amount of the phonon sideband that lies in the cavity linewidth should be at most a few percent and can be ignored.

We validate the cavity QED parameters by reproducing the transmission lineshapes at the two detunings in Fig.~\ref{fig:fig_3}D-E. We find that a nearly lifetime-limited pure dephasing rate of $\gamma^*=10$~MHz gives the best fit for the resonant and dispersive lineshapes, giving cavity QED parameters of $(g, \kappa, \gamma)/2\pi = (0.6, 45, 0.06)$ GHz and a cooperativity of $C=0.53$. 

Another important metric is the proportion of decay into the cavity mode $\beta = \Gamma_{1D}/(\Gamma_{1D} + \Gamma') $, which we determine to be 44\%. The efficient emitter-cavity coupling and low dephasing demonstrate that this platform is appropriate for the development of a push-button photon source.


\begin{figure*}[tb!] 
	\centering
	\includegraphics[width=0.9\textwidth]{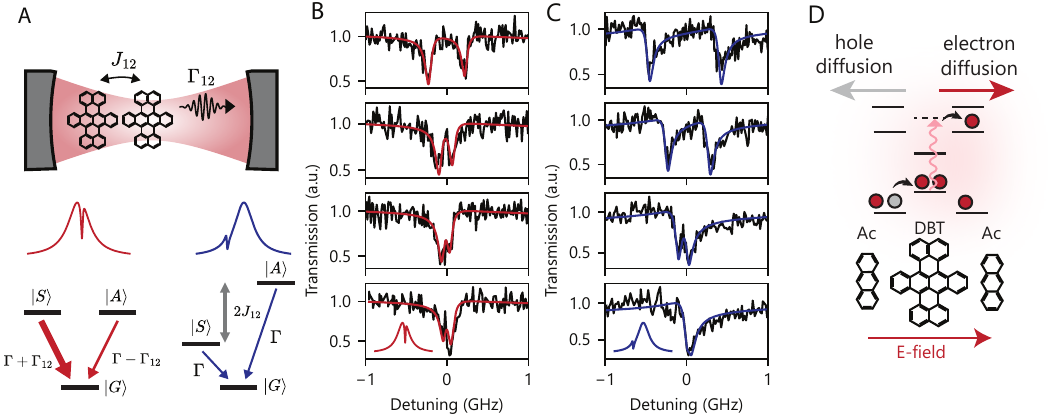} 
\caption{\textbf{Tuning molecules into resonance within a cavity.}
  (\textbf{A}) Two molecules coupled to a photonic crystal cavity can interact through two regimes: dissipatively when resonant to the cavity through the collective decay rate $\Gamma_{12}$ (red, bottom), and dispersively when detuned from the cavity through the dipole-dipole coupling $J_{12}$ (blue, bottom). The energy level diagram shows the ground state $|G\rangle$ and single-excitation manifold with symmetric $|S\rangle$ and antisymmetric $|A\rangle$ states. 
    (\textbf{B, C}) Transmission spectra showing two molecules being tuned toward resonance with each other within the cavity mode. Each panel displays a different step in the tuning process, with experimental data (black) and theoretical fits (red/blue). The left column shows a pair with coupling strengths $g_1 = 560$~MHz and $g_2 = 480$~MHz with collective parameters $J_{12}=-10$~MHz and $\Gamma_{12}= 39$~MHz when the molecules are on resonance; the right column shows a second pair with $g_1 = 610$~MHz and $g_2 = 650$~MHz with $J_{12}=8.7$~MHz and $\Gamma_{12}= 22$~MHz. Red fits correspond to the cavity-resonant (dissipative) regime, while blue fits show the cavity-detuned (dispersive) regime.
    (\textbf{D}) Mechanism of permanent spectral tuning via optically-induced Stark shift. High-intensity excitation creates charge separation in the anthracene matrix surrounding the DBT molecule, with electron (red) and hole (gray) diffusion creating a persistent electric field that produces a quadratic Stark shift of the molecular resonance.
    }
	\label{fig:fig_4} 
\end{figure*}

While coupling single emitters to optical cavities has been demonstrated across various solid-state platforms, bringing two lifetime-limited emitters into resonance within a shared cavity mode has been achieved in only a few systems~\cite{rattenbacher2023onchip, lukin2023twoemitter, nobakht2024cavity} due to the challenges of inhomogeneous and homogeneous broadening. Here, we tune pairs of molecules into resonance and characterize the collective state (see schematic Fig.~\ref{fig:fig_4}A) in both the resonant (dissipative) and detuned (dispersive) regimes of cavity QED.  This represents the first demonstration of resonant multi-emitter cavity coupling achieved without external magnetic or electric fields, made possible by the combination of high doping density, low inhomogeneous broadening, and an optically-induced Stark shift. 

When DBT in anthracene is exposed to high laser power, it undergoes a long-lived frequency shift, shown in Fig.~\ref{fig:fig_4}D.  This mechanism involves electron transfer between the molecule and the surrounding matrix, creating a charge-hole pair in the matrix that produces a DC Stark shift on the DBT molecular resonance~\cite{colautti2020laser, duquennoy2024enhanced}.  This technique results in permanent shifts up to 100 GHz and has been used to create superradiant and subradiant states of two molecules separated by tens of nanometers~\cite{lange2022excitation} and to tune two molecules into resonance a free-space fiber cavity~\cite{nobakht2024cavity}.
As different molecules shift at different rates, we observed molecules crossing each other within the cavity linewidth. To characterize their collective interactions, we fitted their antiresonant lineshapes at large detunings, where interactions were negligible, to extract the individual coupling strengths. Assuming decoherence rates of $\gamma=60$ MHz, similar to the molecule in Fig.~\ref{fig:fig_3}, we extracted values of the cavity emitter coupling rates. We predicted the lineshapes as the molecules approached resonance, finding good agreement with measurements (see Fig.~\ref{fig:fig_4}B-C. 

The collective systems can be characterized by the spin exchange rate $J_{12}$ and the collective decay rate $\Gamma_{12}$
\begin{equation}
    J_{12} \,=\, -\,\frac{g_1\,g_2\,\Delta_{mc}}
  {\Delta_{mc}^2 + \bigl(\kappa/2\bigr)^2},
\qquad
\Gamma_{12} \,=\, \,\frac{g_1\,g_2\,\kappa}
         {\Delta_{mc}^2 + \bigl(\kappa/2\bigr)^2}\,.
\end{equation}
For the pair of molecules in Fig.~\ref{fig:fig_4}(B), we measured with $g_1/2\pi  = 560$~MHz, $g_2/2\pi = 480$~MHz, and a molecule-cavity detuning of $\Delta_{mc}=5.7$~GHz, giving $J_{12}=-10$~MHz and $\Gamma_{12}= 39$~MHz. For the pair of molecules in Fig.~\ref{fig:fig_4}(C), we measured $g_1/2\pi  = 610$~MHz, $g_2/2\pi = 650$~MHz, and $\Delta_{mc}=-17$~GHz, giving $J_{12}=8.7$~MHz and $\Gamma_{12}= 22$~MHz. Coupling multiple emitters to a shared photonic mode opens pathways to several quantum applications, including the generation of squeezed light~\cite{haakh_squeezed_2015}, cluster state preparation~\cite{economou}, and dissipative entanglement generation~\cite{vivas2023dissipative}.

We have demonstrated a hybrid cavity QED platform that achieves strong light-matter coupling between multiple coherent molecular emitters through a single photonic crystal cavity. The key breakthrough lies in integrating the exceptional optical coherence of organic molecules with high quality factor nanophotonics in an approach that scales readily to many-body collective systems and many systems on a chip. 

The potential of this platform can be unlocked with increased cavity quality factors~\cite{zhan2020highq}, which would enable unity light-matter coupling efficiency and approach the strong coupling regime necessary for deterministic single-photon gates and photon-photon interactions. Electrode arrays for spectral tuning~\cite{duquennoy2024enhanced} combined with charge engineering techniques already established in other solid state systems would provide deterministic control over large ensembles of emitters and suppression of residual decoherence. Many-body systems could generate macroscopic superposition states and bright sources of non-classical light with enhanced brightness and reduced noise compared to single-emitter systems.

This work has implications beyond cavity QED as it establishes the viability of chemically synthesized qubits in quantum photonics. This paradigm opens a new design space where quantum properties are engineered through molecular chemistry rather than materials physics constraints. The platform's accessibility and scalability position it to accelerate both fundamental studies of many-body quantum optics and the development of practical applications, including quantum light sources, quantum sensors, and photonic quantum information processors.

\clearpage 


	


\clearpage 

%
\bibliography{main} 
\bibliographystyle{apsrev4-2}

%
%
%
%
%
%


\section*{Acknowledgments}
The authors thank Xue-Wen Chen for assistance with micropositioning techniques and Alex Clark for helpful discussions and comments on the manuscript. 
\paragraph*{Funding:}
This work is supported an Energy Frontier Research Center funded by the U.S. Department of Energy (DOE), Office of Science, Basic Energy Sciences (BES), under award DE-SC0025620. Additional support was provided by the Center for Quantum Technologies under the Industry-University Cooperative Research Center Program at the US National Science Foundation under Grant No. 2224960.

\newpage
\onecolumngrid 

\newcommand{\ket}[1]{\left|#1\right\rangle}
\renewcommand{\thefigure}{S\arabic{figure}}
\renewcommand{\thetable}{S\arabic{table}}
\renewcommand{\theequation}{S\arabic{equation}}
\renewcommand{\thepage}{S\arabic{page}}
\setcounter{figure}{0}
\setcounter{table}{0}
\setcounter{equation}{0}
\setcounter{page}{1} 

\begin{center}
\section*{Supplementary Materials for\\ \scititle}

Christian~M.~Lange,
Arya~D.~Keni,
Ishita~Agarwal,
Emma~Daggett,
Adhyyan~S.~Mansukhani,\\
Ankit~Kundu,
Benjamin~Cerjan,
Libai~Huang,
and Jonathan~D.~Hood$^{\ast}$\\
\small$^\ast$Corresponding author. Email: hoodjd@purdue.edu\\
\end{center}




\subsection*{Materials and Methods}

\subsubsection*{Crystal Synthesis}

DBT-doped anthracene crystals were synthesized using a homebuilt apparatus. A mixture of 15 mg anthracene (99\% purity, Sigma Aldrich) and 15~\textmu L of $10^{-4}$ M DBT in toluene (Symeres) was heated to $250^{\circ}$~C in a glass crucible wrapped in aluminum foil under a nitrogen atmosphere at 550 mbar, with pressure stabilized by a mass flow controller (Alicat). During heating, microdroplets of the molten solution formed airborne crystals that settled throughout the chamber. Crystals were collected on polyvinyl chloride (PVC) plates due to their weak adhesion, allowing pickup via tapered optical fibers. We maximized the yield of high-quality, high-aspect-ratio crystals with two strategies. First, a magnetic shutter (neodymium magnet wrapped in aluminum foil) was placed over the crucible and removed only upon reaching $250^{\circ}$C, preventing formation of low-quality crystals at intermediate temperatures. Second, noting that the high-aspect-ratio crystals remained airborne longer than very small crystals (which fly upward to the roof of the chamber) or very large crystals (which fall under the influence of gravity), PVC collection plates ($\sim$3 cm wide) were positioned in a geometrically protected region accessible primarily through gas flow. Three PVC plates were stacked and separated by spacers, with the bottom plate yielding the highest concentration of pure crystals due to its protection from gravitational debris.

\subsubsection*{Micropositioning}

Crystals were micropositioned using tapered optical fibers attached to a 3D micropositioning stage. Optical fiber segments (Thorlabs 630HP) were hand-tapered by applying axial tension while heating with a flame until the fiber separated. The tapered fiber was mounted on the micropositioning stage and used to pick up crystals via van der Waals forces from polyvinyl chloride (PVC) substrates while being viewed through an optical microscope. The angle between the fiber and the substrate is a critical parameter for positioning, and it should be nearly parallel to the substrate for maximal adhesion. Crystals were stamped onto photonic structures with precise rotational alignment of the crystal's b-axis parallel to the cavity's TE mode polarization, verified under an optical microscope. Following crystal placement, polyvinyl alcohol (PVA) was spin-coated to prevent anthracene sublimation. The silicon nitride rings encircling each cavity served three functions: (1) preventing amorphous PVA from contacting the cavity mode, (2) providing mechanical support to prevent crystal fracture from excessive bending around nanostructures, and (3) ensuring the crystal remained flat for optimal optical contact with the cavity evanescent field. Alternative support structures, such as rails adjacent to the cavity, caused the crystal to bend around the cavity.

\subsubsection*{Cavity Design}

The photonic crystal cavities were designed according to the recipe given in Ref.~\cite{quan2011deterministic}. The nanobeam cavities were constructed from stoichiometric silicon nitride ($n = 2.0$) on silicon dioxide ($n = 1.45$). Each cavity was designed from a beam 556 nm wide and 300 nm thick with 60 holes: 10 mirror holes on each side with 40 cavity holes between them. The Bragg period was held constant across the entire structure, ranging from 240 nm to 247 nm to set the cavity resonant frequency between 765-785 nm before anthracene crystal placement. The mirror holes had a diameter of 136 nm to maximize mirror strength at the target frequency, with the diameter tapering linearly to 180 nm at the cavity center. This quadratic tapering of the filling fraction produces a nearly Gaussian cavity mode with low radiation losses. The cavity diameters are further optimized to maximize the quality factor using a Nelder-Mead optimizer. For the final cavity design, FDTD simulations predict quality factors of $Q = 25,000$ and mode volumes of V = $2.8\,(\lambda/n)^3$. Fabricated devices typically achieved $Q \approx 10,000$ before emitter integration and up to $Q = 8,600$ after integration.


\subsubsection*{Nanofabrication}

The photonic structures were fabricated from 300 nm thick stoichiometric silicon nitride on 4 \textmu m silicon dioxide (Rogue Valley Microdevices). The fabrication process involved: (1) spin-coating 330 nm of Ma-N 2403 resist with Surpass 3000 primer, (2) electron beam lithography using a JEOL 8100XS system (540 $\mu$C/cm$^{2}$ dose, 2 nA beam current), (3) development in Ma-D 525 developer, and (4) reactive ion etching using a Panasonic IE E620CP etcher with CHF$_{3}$ chemistry (2 nm/s etch rate, 1.1:1 selectivity for Si$_{3}$N$_{4}$:SiO$_{2}$).


\subsubsection*{Optical Measurement}

Optical measurements were performed in a Montana cryostat at 3~K using a titanium sapphire laser with $\sim$10~MHz linewidth. The photonic device was accessed through an objective lens (NA $= 0.6$) positioned above the sample, with device imaging provided by an Andor camera (SOLIS X 8181). Cavity transmission signals were measured using an avalanche photodiode (Excelitas SPCM-900-14-FC). Lifetime measurements employed two electro-optical modulators (Jenoptik AM906b and AM830) configured in series and pulsed with a Stanford Research Systems DG645 combined with an SRD1 fast rise time module to achieve laser switching with a 90\% to 10\% fall time of 800 ps. Fluorescence decay traces were recorded using a Swabian TimeTagger Ultra.




\subsection*{Supplementary Text}

\subsubsection*{Theory for Cavity Collective Interactions}

We model \(N\) two-level emitters with transition frequencies
\(\omega_{j}\)
coupled to a single cavity mode with resonance frequency \(\omega_{c}\). The cavity is probed in transmission. The system Hamiltonian is the Tavis-Cummings Hamiltonian which, in a frame rotating at the probe laser frequency $\omega_L$, reads

\begin{equation}
  H \,=\,
  -\hbar\Delta_{c}\,\hat{a}^{\dagger}\hat{a}
  \,-\,\sum_{i=1}^{N}\hbar\Delta_{i}\,\hat{\sigma}^{\dagger}_{i}\hat{\sigma}_{i}
  \,+\,\sum_{i=1}^{N}\hbar g_{i}
          \bigl(\hat{\sigma}_{i}\hat{a}^{\dagger}+\hat{\sigma}^{\dagger}_{i}\hat{a} \bigr) \,
  \label{eq:H_JC}
\end{equation}
where
\(\Delta_{c}\equiv\omega_L-\omega_{c}\)
and
\(\Delta_{i}\equiv\omega_L-\omega_{i}\)
are detunings of the probe from the bare cavity and
emitter~\(i\), respectively, \(\sigma_{i}\equiv|g\rangle_{i}\!\langle e|_{i}\) and \(\sigma^\dagger_{i}\equiv|e\rangle_{i}\!\langle g|_{i}\) are the atomic lowering and raising operators, $a^\dagger$ and $a$ are the creation and annihilation operators of a photon in the cavity mode and \(g_{i}\) is the single-photon
vacuum Rabi frequency of emitter~\(i\).

The corresponding Lindblad operator which captures all dissipative processes reads:
\begin{align}
\mathcal{L}[\hat{\rho}] 
&= \frac{\Gamma'}{2} \sum_{i=1}^N 
   \mathcal{D}[\hat{\sigma}_i]\hat{\rho}  
 + \frac{\kappa}{2} 
   \mathcal{D}[\hat{a}]\hat{\rho}\, + \frac{\gamma^*}{2} \sum_{i=1}^N \mathcal{D}[\hat{\sigma}^z_i]\hat{\rho}
\label{eq:lindbladian}
\end{align}
where we have defined the Lindblad superoperator:
\begin{equation}
    \mathcal{D}[\hat{L}]\hat{\rho} = 2\hat{L}\hat{\rho}\hat{L}-\hat{L}^\dagger\hat{L}\hat{\rho} - \hat{\rho}\hat{L}^\dagger\hat{L}
\end{equation}
$\Gamma'$ is the decay rate to free space and $\gamma^*$ represents
pure dephasing rate arising from elastic processes such as phonon scattering. When coupled to a feedthrough waveguide, the Heisenberg-Langevin equation for an arbitrary system operator $\hat{c}(t)$ is 

\begin{equation}
    \partial_{t}c=-\frac{{\mathrm{i}}}{\hbar}\left[\hat{c},H_{{\mathrm{sys}}}\right]-\left\{ \left[\hat{c},\hat{a}^{\dagger}\right]\left(\frac{\kappa}{2}\hat{a}+\sqrt{\kappa}\hat{a}_{{\mathrm{in}}}(t)\right)-\left(\frac{\kappa}{2}\hat{a}^{\dagger}+\sqrt{\kappa}\hat{a}_{{\mathrm{in}}}^{\dagger}(t)\right)\left[\hat{c},\hat{a}\right]\right\} .
\end{equation}
where $\hat{a}_{\text{in},\ell}(t)$ represents the input field operator coupled through the left mirror of the cavity satisfying $[\hat{a}_{\text{in}}(t), \hat{a}_{\text{in}}^{\dagger}(t')] = \delta(t-t')$. 
The Heisenberg equations of motion for the field and atomic operators are:
\begin{equation}
    \dot{\hat{a}} = \Bigl(\mathrm{i}\Delta_c\, - \frac{\kappa}{2}\Bigr)\hat{a} \,-\,\mathrm{i}\sum_{i=1}^N g_i\, \hat{\sigma}_i \,+\, \sqrt{\kappa} \hat{a}_{\mathrm{in},\ell}(t)
\label{eq:cav}
\end{equation}
\begin{equation}
    \dot{\hat{\sigma}}_i = \Bigl(\mathrm{i}\Delta_{i} -\frac{\Gamma'}{2} -\gamma^*\Bigr)\,\hat{\sigma}_i + \mathrm{i}\,g_i\,\hat{\sigma}^z_i\,\hat{a}
\label{eq:atom}
\end{equation}
Here $\hat{a}_{\text{in},\ell}(t)$ represents the input field operator coupled through the left mirror of the cavity.
When $\Gamma',g \ll \kappa$,  the cavity field can be adiabatically eliminated.
\begin{equation}
    \dot{\hat{a}=0 \,\rightarrow\, \hat{a}=\frac{\mathrm{i}\sum_{j=1}^N g_j \hat{\sigma_j} \,-\sqrt{\kappa}\,\hat{a}_{\mathrm{in},\ell}(t)}{\mathrm{i}\Delta_c\,-\,\frac{\kappa}{2}} }
\end{equation}
Substituting this back into Eq~\ref{eq:atom},
\begin{equation}
    \dot{\hat{\sigma}}_i = \Bigl(\mathrm{i}\Delta_{i}-\frac{\Gamma'}{2}-\gamma^*\Bigr)\,\hat{\sigma}_i -\mathrm{i}\hat{\sigma}^z_i\,\Bigl[\sum_{j=1}^N(J^{ij}_{\mathrm{1D}}+\mathrm{i}\frac{\Gamma^{ij}_{\mathrm{1D}}}{2})\,\hat{\sigma_j}(t)\Bigr] \,-\,g_i\hat{\sigma}^z_i\,\frac{\sqrt{\kappa}\,\hat{a}_{\mathrm{in},\ell}(t)}{\mathrm{i}\Delta_c\,-\,\frac{\kappa}{2}}
\end{equation}
where we have define the spin exchange and decay rates into the cavity mode 
\begin{equation}
    J_{ij} \,=\, -\,\frac{g_i\,g_j\,\Delta_c}
  {\Delta_c^2 + \bigl(\tfrac{\kappa}{2}\bigr)^2},
\qquad
\Gamma_{ij} \,=\, \,\frac{g_i\,g_j\,\kappa}
         {\Delta_c^2 + \bigl(\tfrac{\kappa}{2}\bigr)^2}\,.
\end{equation}
In the low saturation regime $\langle\hat{\sigma}_z\rangle\approx-1$ and the steady state  from Eq~\ref{eq:atom} and \ref{eq:cav}:
\begin{equation}
    \hat{\sigma}_i = \frac{g_i\hat{a}}{\Delta_{i}+\mathrm{i}\Bigl(\frac{\Gamma'}{2}+\gamma^*\Bigr)}\Rightarrow \hat{a}=\frac{-\mathrm{i}\sqrt{\kappa}\hat{a}_{\text{in},\ell}}{(\Delta_c+\mathrm{i}\frac{\kappa}{2})-\sum_{i=1}^N\frac{g_i^2}{\Delta_i+\mathrm{i}(\frac{\Gamma'}{2}+\gamma^*)}}
\label{eq:cavinout}
\end{equation}
For an asymmetric Fabry–Perot cavity
The relation between the input and output modes~\cite{gardiner1985, rempe2015} in the Heisenberg picture is
\begin{subequations}
\begin{align}
  a_{\mathrm{out},\ell}(t) - a_{\text{in},\ell}(t)
  &= \sqrt{\kappa_{\ell}}\;a(t),   \label{eq:IO_left}
  \\[2pt]
  a_{\mathrm{out},r}(t)   - a_{\text{in},r}(t)
  &= \sqrt{\kappa_{r}}\;a(t).      \label{eq:IO_right}
\end{align}
\end{subequations}
We assume the probe enters solely through the left port,
\(a_{\text{in},r}=0\), so the measurable transmission amplitude
is
\(t(\omega)\equiv
  \langle a_{\mathrm{out},r}\rangle/
  \langle a_{\mathrm{in},\ell}\rangle\) and $\kappa_l=\kappa_r=\kappa/2$.
Using Eq~\ref{eq:cavinout} we obtain the intensity transmission spectrum through the cavity
\begin{equation}
   T(\omega)=\left|\frac{\kappa/2}{(\mathrm{i}\Delta_c-\frac{\kappa}{2})+\sum_{i=1}^N\frac{g_i^2}{\mathrm{i}\Delta_i-(\frac{\Gamma'}{2}+\gamma^*)}}\right|^2.
\end{equation}
For a single atom with $J^{ij}_{\text{1D}}\equiv J_{\text{1D}}$ and $\Gamma^{ij}_{\text{1D}}\equiv \Gamma_{\text{1D}}$
\begin{equation}
    \frac{t(\Delta_i)}{t_0(\Delta_i)}=\frac{\Delta_i+\mathrm{i}(\Gamma'+2\gamma^*)/2}{\Delta_i+J_{\text{1D}}+\mathrm{i}(\Gamma'+2\gamma^*+\Gamma_{\text{1D}})/2}.
\end{equation}


\newpage

\begin{figure}
    \centering
    \includegraphics[width=0.9\textwidth]{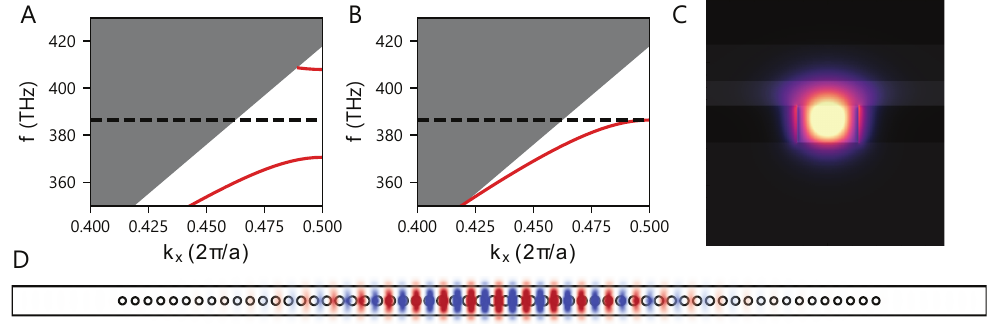}
    \caption{\textbf{Nanobeam cavity design.}
    \textbf{(A,B)} Band structures for the mirror region (A) and cavity center (B) before anthracene crystal and PVA integration. The dashed line at 383 THz (780 nm) shows the mirror region operating within the bandgap while the cavity region operates in the slow-light regime of the dielectric mode. To create the cavity mode, hole diameters are linearly tapered from 136 nm in the mirror region to 180 nm at the cavity center, maintaining constant Bragg periodicity of 245 nm. Hole diameters are further optimized using a Nelder-Mead optimizer to maximize quality factor. The gray region indicates the light line boundary.
    \textbf{(C)} Cross-sectional mode profile of the integrated structure after anthracene crystal integration ($n = 1.8$) and PVA encapsulation ($n = 1.5$). The simulated cavity mode profile is overlaid on the device geometry.
    \textbf{(D)} Electric field profile of the fundamental cavity mode, exhibiting an approximately Gaussian envelope. The simulated quality factor is $Q = 25,000$ with mode volume $V = 2.8(\lambda/n)^3$.
    }
    \label{fig:fig_sm1}
\end{figure}

\begin{figure}
    \centering
    \includegraphics[width=0.9\textwidth]{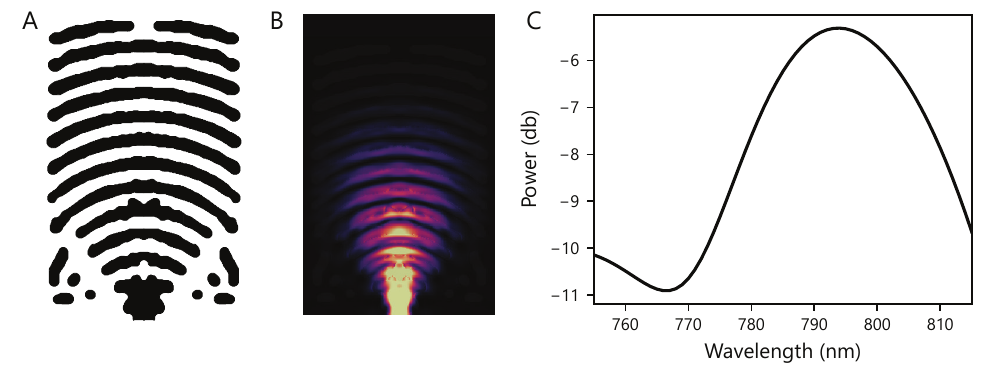}
    \caption{\textbf{Inverse-designed grating coupler optimization.}
    \textbf{(A)} Final grating coupler design after optimization with the Tidy3D adjoint optimizer~\cite{tidy3d}.
    \textbf{(B)} Simulated electric field intensity $|\mathbf{E}|^2$. The grating coupler was designed to couple a Gaussian beam (5~\textmu m waist radius) at $10^{\circ}$ incidence angle to the TE$_{00}$ mode of a ridge waveguide with 500 nm width, 300 nm thickness, and refractive index $n = 2$.
    \textbf{(C)} Broadband coupling efficiency spectrum, achieving peak performance of $-5.3$ dB.
    }
    \label{fig:fig_grating_coupler}
\end{figure}
\begin{figure}
    \centering
    \includegraphics[width=0.9\textwidth]{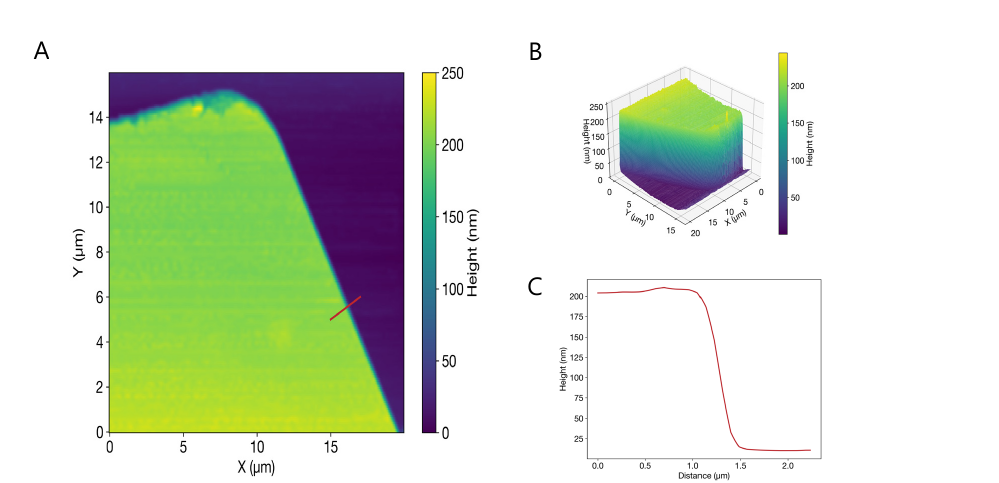}
    \caption{\textbf{Crystal morphology.}
    \textbf{(A)} Atomic force microscopy (AFM) scan of an anthracene crystal.
    \textbf{(B)} Three-dimensional contour profile showing smooth crystal surfaces with minimal defects.
    \textbf{(C)} Cross-sectional height profile confirming crystal thickness of $\sim$200 nm and RMS surface roughness of $\sim$3~nm. }
    \label{fig:fig_sm4}
\end{figure}
\begin{figure}
    \centering
    \includegraphics[width=0.9\textwidth]{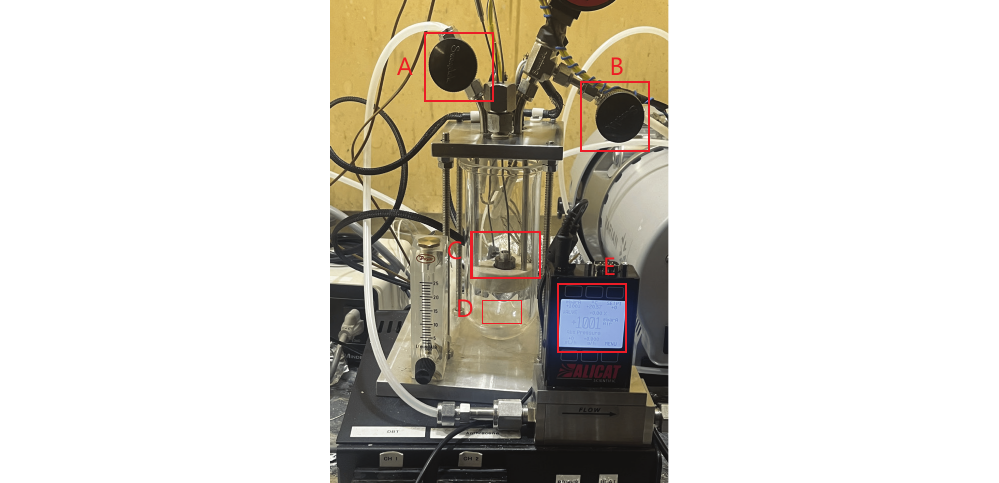}
    \caption{\textbf{Crystal synthesis apparatus.}
    Photograph of the homebuilt synthesis apparatus. \textbf{(A)} Nitrogen gas input. \textbf{(B)} Vacuum output port. \textbf{(C)} Crucible with heating element and temperature probe. A neodymium magnet wrapped in aluminum foil serves as a shutter controlled by an external magnet. \textbf{(D)} PVC collection substrates are positioned in this area beneath the crucible. Three stacked PVC substrates with spacers provide size-selective collection, as high-aspect-ratio crystals remain suspended longer in the nitrogen flow and settle on the bottom substrate, while lower-quality crystals fall onto the upper substrate or adhere to chamber walls. \textbf{(E)} Gas flow is stabilized by an Alicat mass flow controller. In a typical synthesis, 15 mg anthracene and 15~\textmu L of $10^{-4}$ M DBT in toluene are heated to $250^{\circ}$C in a 550 mbar nitrogen atmosphere. After removing the shutter, microdroplets form airborne crystals that settle for a few minutes. The chamber is then flooded with nitrogen and the PVC substrates are removed before the anthracene crystals can sublime.}
    \label{fig:fig_sm8}
\end{figure}


\begin{figure}
    \centering
    \includegraphics[width=0.9\textwidth]{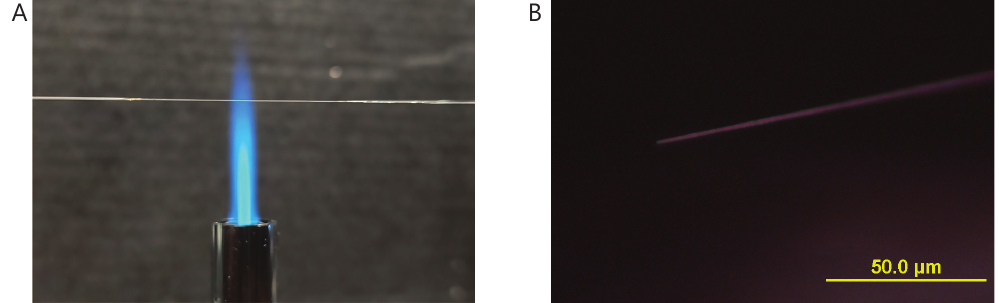}
    \caption{\textbf{Optical fiber tapering.}
    \textbf{(A)} Fiber tapering process. The fiber cladding is first removed by gentle scraping with a razor blade. The exposed core is then heated with a flame while stretching until it breaks. This simple technique can produce fiber tips less than 1~\textmu m for micropositioning anthracene crystals. 
    \textbf{(B)} Optical microscope image of a typical tapered fiber tip.}
    \label{fig:fig_sm10}
\end{figure}





\end{document}